# An Energy Efficient Health Monitoring Approach with Wireless Body Area Networks

SEEMANDHAR JAIN, Indian Institute of Technology Indore, India
PRARTHI JAIN, Indian Institute of Technology Indore, India
PRABHAT K. UPADHYAY, Indian Institute of Technology Indore, India
JULES M. MOUALEU, University of the Witwatersrand, South Africa
ABHISHEK SRIVASTAVA, Indian Institute of Technology Indore, India

Wireless Body Area Networks (WBANs) comprise a network of sensors subcutaneously implanted or placed near the body surface and facilitate continuous monitoring of health parameters of a patient. Research endeavours involving WBAN are directed towards effective transmission of detected parameters to a Local Processing Unit (LPU, usually a mobile device) and analysis of the parameters at the LPU or a back-end cloud. An important concern in WBAN is the lightweight nature of WBAN nodes and the need to conserve their energy. This is especially true for subcutaneously implanted nodes that cannot be recharged or regularly replaced. Work in energy conservation is mostly aimed at optimising the routing of signals to minimise energy expended. In this paper, a simple yet innovative approach to energy conservation and detection of alarming health status is proposed. Energy conservation is ensured through a two-tier approach wherein the first tier eliminates 'uninteresting' health parameter readings at the site of a sensing node and prevents these from being transmitted across the WBAN to the LPU. The second tier of assessment includes a proposed anomaly detection model at the LPU that is capable of identifying anomalies from streaming health parameter readings and indicates an adverse medical condition. In addition to being able to handle streaming data, the model works within the resource-constrained environments of an LPU and eliminates the need of transmitting the data to a back-end cloud, ensuring further energy savings. The anomaly detection capability of the model is validated using data available from the critical care units of hospitals and is shown to be superior to other anomaly detection techniques.

Authors' addresses: Seemandhar Jain, Indian Institute of Technology Indore, Khandwa Road, Indore, India, seemandharj@gmail.com; Prarthi Jain, Indian Institute of Technology Indore, Khandwa Road, Indore, India, prarujain15@gmail.com; Prabhat K. Upadhyay, Indian Institute of Technology Indore, Khandwa Road, Indore, India, pkupadhyay@iiti.ac.in; Jules M. Moualeu, University of the Witwatersrand, Johannesburg, South Africa, Jules.Moualeu@wits.ac.za; Abhishek Srivastava, Indian Institute of Technology Indore, Khandwa Road, Indore, India, asrivastava@iiti.ac.in.

## 1 INTRODUCTION

The world is seeing rapid advancements in the field of 'smart', online health monitoring today. An important contribution towards this is through the effective use of Wireless Body Area Networks, WBANs [9, 28]. A WBAN comprises a network of usually miniaturised computing devices that accommodate independent sensing nodes deployed subcutaneously (just under the skin), over the skin, or over/under clothing depending on the nature of the health parameter being monitored. Each node monitors one or more health parameters and sends the signal(s) back over appropriate communication protocols like the IEEE 802.15.6 [32] across the network of nodes to a Local Processing Unit (LPU) like a mobile device as shown in Figure 1. The LPU collates signals from various sensors and runs analytical algorithms sometimes locally but mostly at a back end cloud to make sense of the health parameter values. This works like a well oiled machine and the well being of an individual is effectively monitored. The issue arises when one or more sensing nodes run out of energy and is unable to send back signals nor facilitate the passage of signals from other nodes. This is an unsurprising eventuality given the small and constrained architecture of a node and is especially true for nodes implanted subcutaneously or deep inside the individual's body with no means to recharge. Several endeavours are directed to address this with a large fraction devoted to optimising the route of signal transmission from the sensing node to the LPU [4, 44, 49, 54]. Some work is in the direction of utilising alternative communication protocols to reduce the energy expended [6, 29, 53].

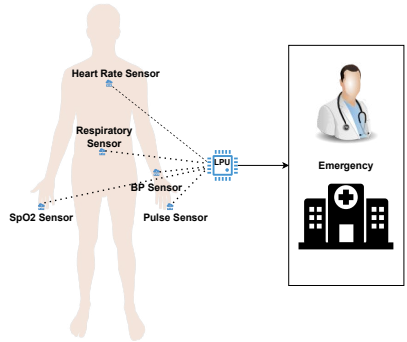

Fig. 1. Representational depiction of WBAN

We propose a simple yet effective approach to energy conservation in this paper. Unlike earlier work that is directed towards optimally transmitting signals to the LPU, our approach conserves energy by 1) reducing the data that needs to be transmitted to the LPU; and 2) eliminating the need to transmit collated signals from the LPU to the back-end cloud for analysis. The first of this two-tier approach involves significant reduction of data that needs to be transmitted to the LPU. This is done by eliminating *uninteresting* and *visibly faulty* data at the site of the sensor node. Uninteresting data here implies health parameter values whose deviation from the immediately preceding value is very small and/or whose value is not part of an apparent trend, such that it does not convey useful information. Visibly faulty data implies health parameter values that are so widely astray that it is safe to assume a fault in their provenance most likely in the functioning of the sensor. Profiling and eliminating data in this manner at the site of the sensor node is non-trivial given the resource-constrained nature of these nodes. We, therefore, employ an algorithm of $O(1)$ time and space complexity for this purpose and demonstrate through hardware simulation the practical viability of the approach.

The second tier of our approach reduces the energy expended by eliminating the need to transmit data to the back-end cloud for analysis. An anomaly detection approach that works effectively with streaming data is utilised without the need to store data. The anomaly detection approach is able to identify anomalies in the health parameters of the monitored individual at the LPU and indicates an adverse health condition.

The proposed two-tier approach is validated on data available in the MIMIC Critical Care Database [1] (comprising data from the critical care units of several hospitals) and includes the following parameters: Heart Rate, Systolic Blood Pressure, Diastolic Blood Pressure, Pulse rate, Respiratory rate, and Oxygen saturation. The broad contributions of this paper include:

- proposal of a novel approach to energy savings in a WBAN by eliminating faulty and uninteresting data at the sensor node and saving on transmission energy;
- the implementation of a light-weight anomaly detection technique at the LPU, thus minimising the need of a back-end cloud;
- validation of the proposed two-tier approach on standard data-sets and through a real-world prototypical implementation.

The remainder of this paper is organized as follows: Section 2 contains information and discussion on related work and surveys. Section 3 provides details of the proposed approach. In Section 4, the efficacy of the approach is validated through experiments, and finally Section 5 concludes the paper.

## 2 RELATED WORK

E-Health Monitoring is emerging as an important area of research. WBANs play a significant role in these and help in effectively monitoring patients [27, 31]. One of the most important challenges in WBANs is to find effective means to prolong the active life of a sensor node. Several researchers propose prolonging the life of a sensor node by proposing approaches to conserve the latter's energy [2, 6, 58].

Another issue with WBANs functioning as effective health monitoring deployments are faulty measurements. Faulty measurements arise from equipment malfunctions, faulty readings, or environmental noise and lead to unnecessary consumption of sensor energy and possible false alarms. Such faulty measurements need to be detected early and eliminated. Major endeavours in this direction include [3, 45, 52, 60]. Conforming to expectations, several endeavours are towards anomaly detection in WBANs to detect medical issues in patients. Key surveys in this direction include [5, 43, 56].

Anomaly detection techniques are classified as statistical-based [57], Machine learning based [47], Game-based [48], and graph-based [16].

Salem *et al.* [47] propose a Support Vector Machine (SVM) based approach for anomaly detection in WBAN. They construct a classification model comprising SVM and linear regression. The incoming data point from the WBAN is collected at the processing unit and is identified as abnormal if it deviates from the developed model. A data point so categorised as abnormal is further assessed to be an anomaly or a fault using linear regression. In addition to this, they propose a 'counter' mechanism to keep track of spatial correlations between attributes. If the *counter* > 1, an alarm is raised indicating an anomaly; otherwise the data point is construed as faulty and discarded. There are several drawbacks of this approach: 1) it is possible that two or more attributes incur faulty measurements and the model predicts it to be an anomaly whereas in reality the measurement is faulty; 2) the model is not adaptive and is thus not suitable and effective in the real-world where data variation is common. Banerjee *et al.* [8] propose an algorithm that assesses an incoming data point as abnormal using minimum and maximum parametric values that are globally set. If the measured data point is beyond the range of the minimum and maximum values, it is considered to be an abnormal data point. The issue with this simple technique though is that the assumed global minimum and maximum values may not work for all patients [51]. Osman *et al.* [46] propose a classification based approach that utilises the *j*48 decision tree algorithm to classify data points as normal or abnormal in WBAN. If a point is abnormal, they further apply linear regression and correlation to further classify the point as anomalous or faulty. The issue with this approach is that there is no attempt to update the model periodically to conform with the manner data varies in the real world. Furthermore, they pre-define values for health parameters that are considered normal which may not hold for all patients.

Qi *et al.* [42] propose a system called RadioSense, that recognizes body sensor network activity using wireless communication patterns. The authors use a smoothing window size of 9 seconds in their work. The work is primarily concerned with energy costs and uses the K-Nearest Neighbor (KNN) algorithm as the classifier. Li *et al.* [34] propose the use of a WSN-WiFi hybrid network that improves the quality of wireless communication in body area networks by considering both WiFi and ZigBee packet sizes. The study demonstrates the critical role of packet size selection in energy conservation in a hybrid WSN-WiFi hybrid network. Huynh *et al.* [24] propose an energy efficient solution in healthcare systems for WBAN based on ZigBee. The authors use OPNET for simulation, and utilize ZigBee in a beacon-enabled mode. This approach is effective in avoiding energy consumption in the idle mode. However, this endeavour fails to eliminate communication overhead. This leads to congestion that ultimately results in significant increase in the average end-to-end latency. The work by Chiang *et al.* [14] focuses on reducing the energy required for transmitting data for seizure detection and for detecting other related medical conditions. The idea is to not transmit the entire EEG signal but only a few distinct features from a few EEG channels for further analysis. This results in significant reduction in the number of features transmitted and hence transmission energy. The approach, however, does not minimize the energy consumption at the sensor. It is also not a realistic approach as medical practitioners prefer seeing the entire EEG rather than just specific features. Majumdar *et al.* [38] propose the use of Blind Compressed Sensing (BCS) to reduce data transmission and minimize energy consumption associated with sensing, computing, and transmission in WBAN. However, they are unable to regulate the upload frequency of body sensors and the approach is thus not effective in practical settings.

Several other techniques for fault and anomaly detection include one that uses Mahalanobis distance [35], another uses kernel density estimation [61], and third chi-square distance [47] to identify anomalies in data. These are effective but suffer from high computational costs that a WBAN deployment can ill afford. Isolation Forest [36], PiForest [25], and RRCF [21] are tree-based anomaly detection approaches that have conservative computational costs but can only be used for anomaly detection. They are not effective with faulty data and give rise to a large number of false alarms.

More specific architectures that detect vital signs [59] of patients over WBANs include AID-N [18], UbiMon [40], Codeblue [39], AlarmNet [55], and CareNet [26].

## 3 PROPOSED APPROACH

We propose a two-tier anomaly detection approach in a WBAN deployment that conserves energy without compromising on effectively monitoring the health parameters of the patient. The first tier comprises interventions at the site of the WBAN node such that the data transmitted over the WBAN to the LPU is reduced. The reduction of data is achieved by eliminating health parameter values that convey little or no new information. These are mostly parameter values that have not deviated much from the immediately preceding reading or are values that are significantly deviated from the expected and hence are assumed to be faulty. Eliminating such data results in large energy savings without compromising on the relevant information sent back for processing. The second tier of the approach comprises endeavours to analyse data for detecting anomalies at the LPU rather than sending it to the back-end cloud. This is non-trivial as an LPU is not rich enough in terms of resources to store health parameter values emanating at quick intervals and rapidly accumulating to large volumes. An anomaly detection technique that works on streaming data is therefore proposed that can effectively analyse health parameters and identify anomalous health conditions at the LPU. Not sending data to the back-end cloud for analysis significantly conserves energy and eliminates latency.

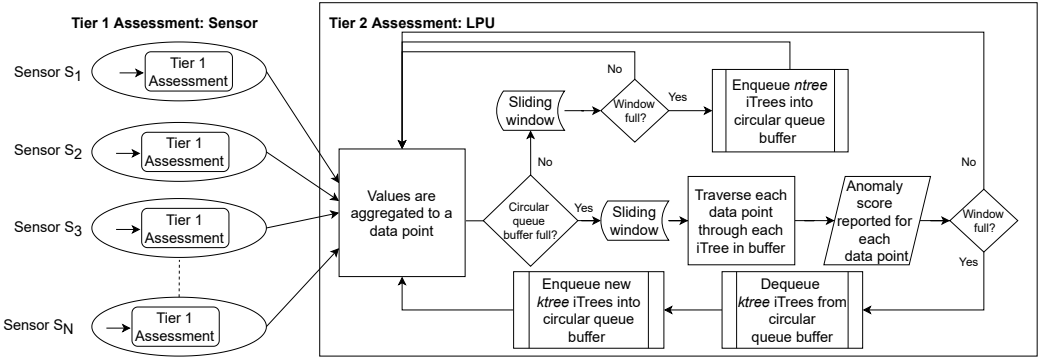

Fig. 2. Proposed Framework

## 3.1 Tier-1 Assessment

Tier-1 implies the site of individual WBAN nodes that comprise one of more sensors. The data on the patient's health originates at these nodes in the form of sensor readings. WBAN nodes are compact, miniaturised devices that accommodate one or more sensors and are often deployed subcutaneously (just below the skin) or much deeper in the patient's body. In some cases, these are deployed above the skin, below or above clothing. In any case, WBAN nodes are extremely resource-constrained both in terms of storage and processing capability. The aim of identifying and eliminating uninteresting and faulty data in such constrained environments is non-trivial. Uninteresting data, as we earlier described, is data that does not deviate much from the immediately preceding data and thus transmitting this data over the WBAN to the LPU does not provide new information. Faulty data comprises health parameter values that are unrealistically removed from normal levels and originate from damaged sensors, incorrect readings, environment noise, and other such issues.

Assuming there are $N$ sensors deployed in the body of a patient as shown in Figure 1. Each sensor senses a health parameter that has one or more attributes (in general $K_i$ attributes such that $K_i \geq 1$), for instance, Blood Pressure has 3 attributes: Systolic Blood Pressure, Diastolic Blood Pressure, and Mean Blood Pressure. The data generated at each time interval by a sensor $S_i$ is $x_{ijt}$ ($i \in [1, N]$ and $j \in [1, K_i]$) where $x_{ijt}$ is the reading of the $j^{\text{th}}$ attribute of the parameter sensed by sensor $S_i$ at the $t^{\text{th}}$ second. The readings of $N$ sensors at time interval $t$ can be visualized as a list such that each entry represents a sensor reading:

$$X_t = \begin{bmatrix} y(1)_t & y(2)_t & y(3)_t & \ldots & y(N)_t \end{bmatrix} \quad (1)$$

where $y(i)_t = \begin{bmatrix} x_{i1t} & x_{i2t} & x_{i3t} & \ldots & x_{ik_it} \end{bmatrix}$.

The $i^{th}$ sensor senses $k_i$ attributes of a health parameter and these have $k_i$ different data values. At time $t$, the data values generated by the $i^{th}$ sensor is represented as $y(i)_t$. The value of each health parameter attribute measured by a sensor undergoes an assessment at each time interval at the site of the node that we refer to as Tier-1 assessment (as shown in Figure 2). During the Tier-1 assessment, the mean ($m$), variance ($v$), and the number of data values generated and measured ($n$) so far are computed. $n$ is updated every $M$ hours (in our experiments, we use $M = 2$ hours because it is long enough to properly study the working of the system and short enough to practically conduct experiments).

For each health parameter attribute measured by sensor $S_i$, a $Z$ value is calculated as:

$$Z = \frac{(x_{ijn} - m)}{\sqrt{v}} \tag{2}$$

where $m$ is the mean of the last $(n-1)$ attribute values and $v$ measures how much variance the attribute values have from the mean value. In general, the mean and variance for any list of $A$ values is calculated as:

$$m = \frac{\sum_{i=1}^{n} A_i}{n}, \tag{3}$$

$$v = \frac{\sum_{i=1}^{n} (m - A_i)^2}{n} \text{ or } v = \frac{\sum_{i=1}^{n} A_i^2}{n} - m^2 \tag{4}$$

If the $Z$ value, so calculated, falls outside the range of $l_{th}$ and $h_{th}$ ($Z < l_{th} || Z > h_{th}$), where $l_{th}$ and $h_{th}$ are the lower and upper thresholds respectively (these are globally declared separately for each health parameter attribute based on universally accepted practices), then the attribute value is discarded as faulty. Otherwise, $\Delta Z$ for the attribute is calculated as:

$$\Delta Z = |(Z - Z_{prev})|, \tag{5}$$

where $Z_{prev}$ is the value of $Z$ at the immediately preceding time interval $t-1$.

The thresholds are calculated based on discussions on the range of vital signs in seminal works in the field [13, 15, 19, 37]. Lockwood *et al.* [37] and Chester *et al.* [13] analyse extensively the existing literature to converge on a consensual range for vital signs across categories of patients such as children, adults, the elderly, neonatal, paediatric, and hospitalised adults. Edmonds *et al.* [15] and Genes *et al.* [19] in their work put together a table that shows the standard deviations and range of vital signs. Using these detailed studies, there is a consensus in the community on a universal set of values for body vitals. We utilise these values to calculate $l_{th}$ and $h_{th}$ for each health parameter attribute.

If $\Delta Z < \epsilon$, (where $\epsilon$ is a small value, $\epsilon = 0.2$ in our experiments) for an attribute value then that attribute value is deemed *uninteresting* for not having deviated much from the immediately preceding value. This uninteresting point is eliminated and not transmitted to the LPU to conserve energy. If the LPU does not receive an attribute value at a particular time interval, it assumes the attribute value to be unchanged from the value at the immediately preceding time interval. This simple procedure is described in Algorithm 1. The mean and variance values for each attribute are updated at each time interval as described respectively:

$$m = \frac{nm + x_{ij(t+1)}}{n+1}, \tag{6}$$

$$v = \frac{n}{n+1} \left( v + \frac{(m - x_{ij(t+1)})^2}{n+1} \right) \tag{7}$$

The Tier-1 assessment process has a time complexity of $O(1)$ time and its space complexity is $O(k)$ where $k = \max(K_1, K_2, \ldots, K_N)$. $K_i$ is the number of health parameter attributes measured by sensor $i$ and is bounded by a constant value [20]. The space complexity, therefore, is effectively $O(1)$. A constant time and space complexity ensures that the Tier-1 assessment is seamlessly conducted within the resource-constrained confines of a WBAN node.

---
**Algorithm 1** Tier-1 Assessment
---
**Input:** Sensor $S_i$ with attribute $j$, current data value $x_{ijt}$
**Output:** $x_{ijt}$ or Discard $x_{ijt}$
1: $Z_{new} = \frac{(x_{ijt} - m)}{\sqrt{v}}$
2: **if** ($Z_{new} \leq h_{th}$ and $Z_{new} \geq l_{th}$) **then**
3: $\quad \Delta Z = abs(Z_{new} - Z_{old})$
4: $\quad$ **if** $\Delta Z \geq \epsilon$ **then**
5: $\quad\quad m = \frac{(n-1)m + x_{ijt}}{n}$
6: $\quad\quad v = \frac{n-1}{n}(v + \frac{(m - x_{ijt})^2}{n})$
7: $\quad\quad$ Transmit $x_{ijt}$ to LPU
8: $\quad$ **else**
9: $\quad\quad$ Discard $x_{ijt}$ as a uninteresting point
10: $\quad\quad$ Send *ACK* bit periodically after a fixed interval of time
11: $\quad$ **end if**
12: **else**
13: $\quad$ Discard $x_{ijt}$ as a faulty point
14: **end if**
15: $Z_{old} = Z_{new}$

---

## 3.2 Tier-2 Assessment

The attribute values of various parameters are transmitted to an LPU and assessment of this data at the LPU comprises Tier-2 Assessment. Given the relative resource restrictions of an LPU, the norm is to further transmit the data received to a back-end cloud for analysis and to draw conclusions on the well being of a patient. We propose an approach to appropriately analyse data at the LPU without the need to send it to the cloud. This has the potential to ensure significant savings in terms of transmission energy as well as reduced latency. It may, however, sometimes be necessary to send data to the cloud. This could be data sent to facilitate more detailed analysis as well as to maintain a history of the recorded vitals. The proposed approach facilitates such kind of interaction and places no limitation on sending data to the cloud. However, to cater for the immediate application, interaction with the cloud is not essential. The issue with analysis of WBAN data at an LPU is that data (in the form of attribute values for various health parameters) arrives at regular intervals and accumulates to quickly take up large volumes. Storing such large volumes of data at the LPU for analysis is non-trivial. We propose an approach that is capable of detecting anomalies in health parameter values that are 'streaming in'. In other words, the proposed approach eliminates the need to store the health parameter values for analysis, rather it analyses the data as it comes in and extracts relevant information from it.

We utilise the *iForest* algorithm for anomaly detection [36] and appropriately amend its usage to work with streaming data within an LPU. *iForest* is a well-known tree based anomaly detection algorithm that is based on randomization [10]. Randomization is a powerful tool used to take out random sub-samples from the dataset. In randomization, an attribute is randomly selected and then an attribute value is randomly chosen from the selected attribute. This value is set for partitioning the root node. The data points whose values are smaller than the selected split value are taken in the left subtree and the data points whose values are greater than the selected split value are taken in the right subtree. Each of the two subtrees is further partitioned in the same way until all the data points are isolated or the maximum tree height is reached, which forms the leaves of the tree. This forms an *iTree*. Multiple *iTrees* are created to form an *iForest*. The point that isolates closer to the root node is an anomaly and one that isolates far from the root node is a normal point.

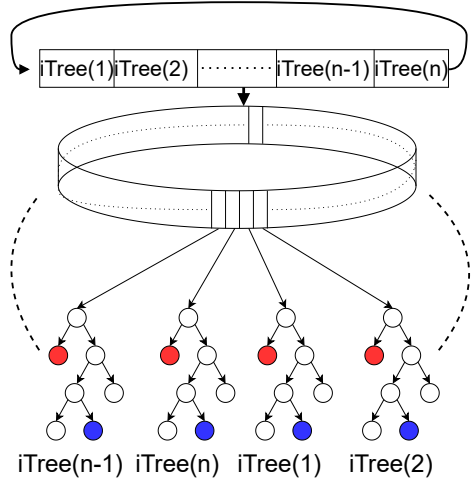

Fig. 3. Circular queue buffer representation

In the proposed approach, attribute values of health parameters across the body converge at the LPU. These attribute values are aggregated to form a single data vector. Each attribute value is normalised and becomes one dimension of the vector. As per the nature of WBAN networks, the attribute values stream in to the LPU and these are handled as follows:

**Sliding Window:** As the attribute values stream in to the LPU, they are quickly aggregated to a single data vector such that one vector corresponds to one time interval. These data vectors are made to fill up a sliding window which is a mechanism useful for performing computations on streaming data. A sliding window allows continuous streaming data to collect in it and cordons off blocks of data of a fixed size ($\omega$). These blocks of $\omega$ data elements are considered for computation one at a time. After these computations are done, the data elements are eliminated and a new set of $\omega$ data elements are considered.

**Creation of *iForest*:** The $\omega$ data vectors in the sliding window are used to create *ntree* number of *iTrees* following the process described in [36] wherein a forest of *iTrees* called *iForest* is created. The process of creating an *iForest* was briefly described at the beginning of this sub-section.

**Populating a Circular Queue Buffer:** The *iTrees* so created are made to populate a buffer in the form of a circular queue wherein each element of the queue has a pointer pointing to the root node of an *iTree*. There are *ntree* elements in the circular queue corresponding to the number of *iTrees* in the *iForest* created. Figure 3 is a representational depiction of a circular queue buffer.

**Identifying anomalous data points:** The data vectors in the sliding window are made to traverse each of the *iTrees* in the circular queue buffer and depending on how quickly the data vector isolates in each case, an assessment is made on whether it is an anomaly or not. The following approach borrowed from the concept of Binary Search Trees (BST) is adopted for calculating the anomaly score.

To compute the anomaly score, a data point is traversed from the root and terminates at an external node i.e., the height of the *iTree* gives an idea of the chance of the data point being an anomaly or not. As the number of nodes in an *iTree* increases at the rate of $O(n)$, the average height of the *iTree* increases at the rate of $O(\log n)$ as *iTrees* have a structure similar to BSTs. To compare the path length until a data point reaches an external node, and effectively the anomaly score, a normalized average height of *iTrees* is computed (as there are several *iTrees* of different heights). This gives a normalized anomaly score. The calculation of the average height of an *iTree* is borrowed from the calculation of unsuccessful search in a BST as

$$d(n) = 2H_{n-1} - (2(n-1)/n) \tag{8}$$

where $n$ is the number of instances in the sample, $H_{n-1} = \ln(n-1) + \gamma$, and $\gamma$ is Euler's constant with a value of approx 0.5772156. As $d(n)$ is the average of $h(n)$, we use it to normalise $h(n)$ [36].

$$S(\mathrm{x}, \omega) = 2^{-E(h(x))/d(\omega)} \tag{9}$$

where $S(\mathrm{x}, \omega)$ is the score that indicates whether a point is an anomaly or not, $\omega$ is the sliding window size, $h(x)$ denotes the length of an *iTree*, $E(h(x))$ denotes the average of $h(x)$ from all the *iTrees* created and $d(\omega)$ is the average of $h(x)$ considering $\omega$ data elements in the sliding window. If $S(\mathrm{x}, \omega) > 0.5$, the data point is an anomaly otherwise it is a normal point.

**Refreshing the Circular Queue Buffer:** After all the data vectors in the sliding window are done traversing the *iTrees* in the buffer, *ktree* number of *iTrees* (where *ktree* < *ntree*) are removed from the circular queue and these *ktree* elements are re-populated with the next batch of *iTrees* from the fresh data in the sliding window. Partially refreshing the circular queue buffer reduces bias in computing the anomaly score.

**Refreshing the Sliding Window:** The data in the sliding window is replaced with a fresh batch of $\omega$ data vectors. These are used to create *ktree* number of *iTrees* and these populate the circular queue buffer.

Anomalous health parameters are so detected in the proposed approach. Subsequent to detection of an anomaly, the LPU can be programmed to store and report the health parameter values responsible for the anomaly to a practitioner and/or raise an alarm and call for help.

Algorithm 2 discusses the step-wise procedure followed in Tier-2 at the LPU to identify anomalous health parameter values in the patient.

## 4 EXPERIMENTAL VALIDATION

To assess the efficacy of the proposed two-tier approach to monitor the well being of a patient whilst simultaneously conserving energy, we utilise a dataset from the MIMIC database [1] that comprises freely available critical care data. We specifically work with the data of two patients: one with id number 221 and the other with id number 230. We choose these two patients because the models with which we compare our approach of anomaly detection have used the data of these two patients to demonstrate their working and hence it is easier to compare. We first demonstrate the efficacy of our approach to assess faulty parameter readings at individual sensor sites. We also demonstrate how a large fraction of sensor readings are uninteresting and compute the percentage of energy

---

**Algorithm 2** Tier-2 Assessment

---

**Input:** $ntree$ - Number of trees, Streaming input data $X = \{X_1, X_2, \ldots, X_i, \ldots, \}$, $\omega-$ Sliding Window Size, $ktree$-number of trees to update $Q$

**Output:** Anomaly Detector

1: *Initialize Circular_Queue_Buffer $F \leftarrow []$*
2: *Initialize Sliding Window as $Y \leftarrow []$*
3: *Initialize count of Streaming data point in window c=0*
4: *while c != $\omega$ do*
    *Insert $X_i$ in Y*
    $c \leftarrow c + 1$
5: *end while*
6: *Apply preprocessing step to Y*
7: *Initialize starting pointer of a circular queue $s \leftarrow 0$*
8: *Initialize height $h \leftarrow ceil\ (log_2(\omega))$*
9: *for $i \leftarrow 0$ to ntree do*
    $F[i] \leftarrow iTree(Y,0,h)$
10: *end for*
11: *Re-initialize $Y \leftarrow []$ and $c \leftarrow \omega$*
12: *while c > 0 do*
    *Insert $X_i$ in Y*
    $c \leftarrow c - 1$
13: *end while*
14: *Apply preprocessing step to Y*
15: *report anomaly detector $G(Y)$*
16: *for $i \leftarrow s$ to $s + ktree$ do*
    *set $F[i\ modulo\ ntree] \leftarrow NULL$*
17: *end for*
18: *for $i \leftarrow 0$ to ktree do*
    $F[s] \leftarrow iTree(Y,0,h)$
    $s \leftarrow s+1$
    $s=s\ modulo\ ntree$
19: *end for*
20: *goto 11 for upcoming data points*

---

savings ensured by eliminating these. Subsequently, Tier-2 efficacy is demonstrated by correct identification of anomalous health conditions taking all parameters into account simultaneously at the LPU. The working of our approach is compared and shown to be superior to the classical *iForest* [36] and the Robust Random Cut Forest (RRCF) based anomaly detection approach [21]. In addition to demonstrating the efficacy of the algorithms at the two tiers, we simulate the hardware of a typical WBAN node using MATLAB/Simulink and run the Tier-1 algorithm over this simulation. The algorithm works exactly as expected validating not just its efficacy but also demonstrating that the Tier-1 set-up can be effectively implemented in the resource-constrained environment of a WBAN node.

The experiments are conducted on a personal PC with Intel(R) Core(TM) i7-7500U CPU @2.70GHz 2.90 GHz and 16 GB memory (RAM). The operating system is Windows 10 Pro. The algorithms are programmed with *Python* 3.7.

## 4.1 Dataset

The dataset, as stated earlier, is part of the MIMIC database [1] and we specifically utilise the data corresponding to two patients with ids 221 and 230. The data of these patients have been used by competing methods. The following six health parameters are monitored for these two patients: 1) Heart Rate; 2) Systolic Blood Pressure; 3) Diastolic Blood Pressure; 4) Pulse rate; 5) Respiratory rate; and 6) Oxygen saturation (SpO2). The data is collected over a duration of 6 hours and 56 minutes from 11:50:32-18:47:12 dated 18/05/1995 and 12:43:50-19:40:14 dated 20/06/1995 for patients 221 and 230 respectively. Table 1 is a description of the data collected for these two patients. Anomalies in the readings are injected synthetically and the percentage of anomalies injected is shown in the table under 'Anomaly Threshold'.

Table 1. Data of patients 221 and 230

| Patient id | Number of Records | Monitoring Duration | Number of Attributes | Anomaly Threshold |
|---|---|---|---|---|
| 221 | 25,000 | 6.94 Hrs | 6 | 6.46% |
| 230 | 25,000 | 6.94 Hrs | 6 | 3.02% |

The six health parameters for patient 221 are shown in Figure 4 and for 230 in Fig. 5.

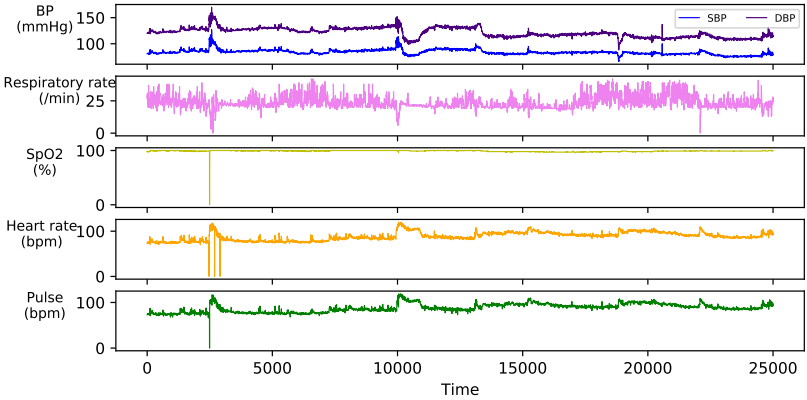

Fig. 4. Patient 221 dataset

## 4.2 Tier-1 Assessment Results

Tier-1 assessment is the assessment of health parameter values at the site of the sensor immediately after they have been generated and before they are transmitted over the WBAN to the LPU. Tier-1 assessment happens at different sensor locations for each of the following health parameters in our experiments: 1) Heart Rate; 2) Systolic Blood Pressure; 3) Diastolic Blood Pressure; 4) Pulse rate; 5) Respiratory rate; and 6) Oxygen saturation (SpO2). The assessment is done to identify health parameter values with one of the following two characteristics:

**Faulty parameter readings:** These are health parameter values that arise out of faulty measurements, malfunctioning equipment, and other such issues. Such measurement are far removed from the normal values such that they fall beyond a range effectively computed by the Median

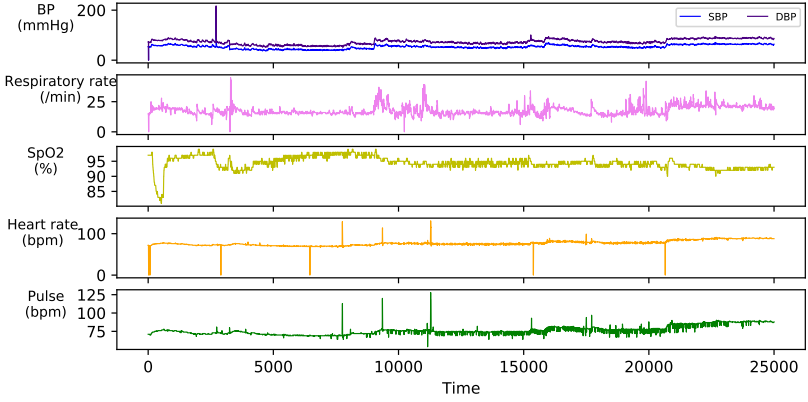

Fig. 5. Patient 230 dataset

Absolute Deviation (MAD) for the data [33]. Our Tier-1 assessment is effective in identifying faults in the various health parameter values and these are shown with black lines in Figures 6 and 7, respectively for patients 221 and 230. The faulty readings so identified are eliminated at the site of the respective sensor and not transmitted over the WBAN to the LPU.

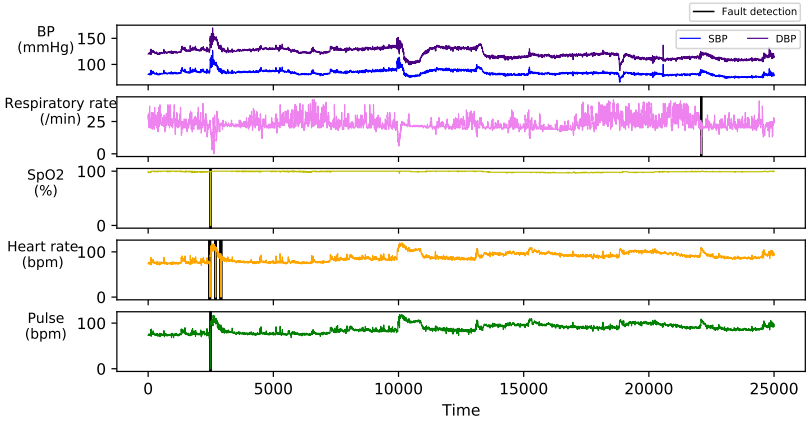

Fig. 6. Fault Detection for Patient 221

**Uninteresting readings:** The other set of health parameter values identified at Tier-1 are *uninteresting readings*. These are parameter values that very slightly deviate from the immediately preceding value such that their transmission over the WBAN does not provide useful information. A significant number of readings are found to be uninteresting and these are eliminated at the site of the sensor node instead of transmitting them over the WBAN to the LPU. The LPU on not receiving a reading at a time interval, assumes it to be uninteresting or in a small number of cases faulty, and uses the value of the immediately preceding time interval for its assessment. Uninteresting readings are assessed if the deviation of the $Z$ value of a reading (calculation of $Z$ is discussed in detail in Section 3.1) is less than a very small value ($\epsilon$) from that of the immediately preceding reading. We experiment with several values of $\epsilon$ between 0 and 1 and the percentage of *uninteresting* health parameter values eliminated is shown in Figure 8. The graph clearly shows that even with a very small value of $\epsilon$, a very large percentage of data points is uninteresting and is eliminated. This results in significant savings of energy that would otherwise have been spent on transmitting this data through the WBAN to the LPU. For a value of $\epsilon = 0.2$, Table 2 shows the actual percentage of health parameter readings discarded for both patients, 221 and 230. An average of around 88% of readings for patient 221 and 93% for patient 230 across the various health parameter sensors are eliminated leading to significant conservation of energy without compromising on monitoring accuracy.

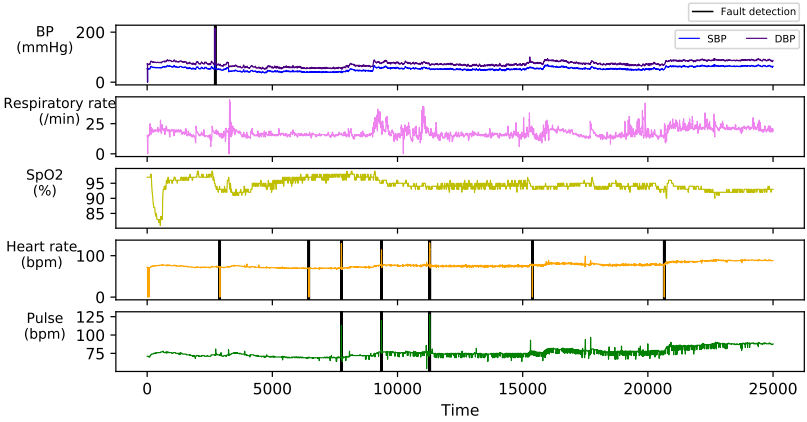

Fig. 7. Fault Detection for Patient 230

Table 2. Outcome of Tier-1 Assessment

| Patient id | | RESP | BP-S | BP-D | SpO2 | HR | PULSE | Total |
|---|---|---|---|---|---|---|---|---|
| **221** | Total # of readings | 25,000 | 25,000 | 25,000 | 25,000 | 25,000 | 25,000 | 1,50,000 |
| | # of discarded uninteresting readings | 16,976 | 19,826 | 23,613 | 24,430 | 23,789 | 23,410 | 1,32,044 |
| | % of readings discarded | 67.904 | 79.304 | 94.452 | 97.72 | 95.156 | 93.64 | **88.029** (average) |
| **230** | Total #of readings | 25,000 | 25,000 | 25,000 | 25,000 | 25,000 | 25,000 | 1,50,000 |
| | # of discarded uninteresting readings | 21,282 | 24,585 | 24,547 | 24,040 | 22,658 | 23,418 | 1,40,530 |
| | % of readings discarded | 85.128 | 98.34 | 98.188 | 96.16 | 90.632 | 93.672 | **93.68** (average) |

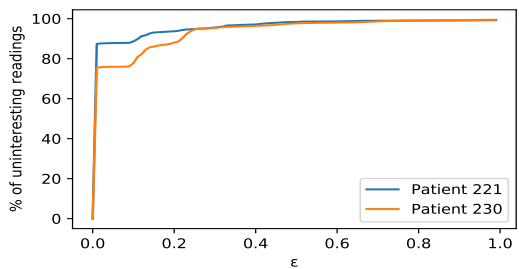

Fig. 8. Percentage of uninteresting readings

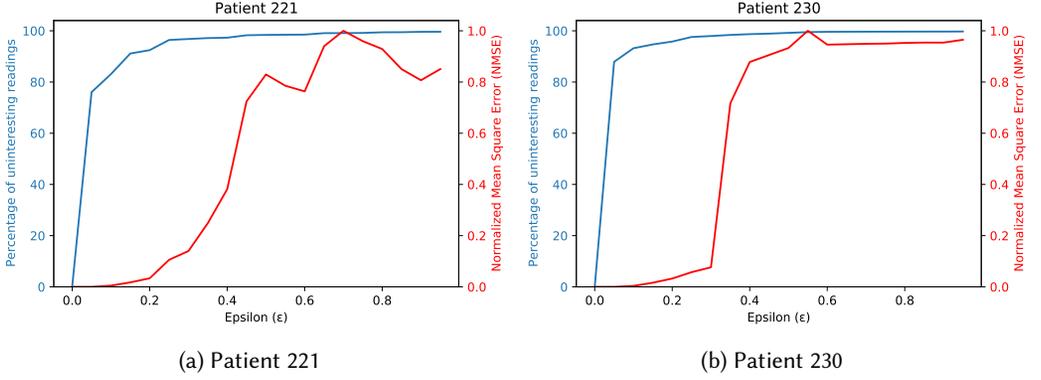

(a) Patient 221      (b) Patient 230

Fig. 9. Graph for trade-off between Percentage of uninteresting readings and NMSE based on different values of epsilon ($\epsilon$)

For the best trade-off between data elimination and information loss, we study the variations in Normalized Mean Square Error (NMSE) and percentage of uninteresting readings with Epsilon. Mean Squared Error gives the average of the square of the difference between the actual and predicted values, and to eliminate biasing of the metric towards the model, the value is normalised between 0 and 1 and this is called normalised MSE.

To find the optimal value of $\epsilon$, NMSE is computed between the original dataset and the output of Tier-1 Assessment. Figure 9 shows the variations of NMSE and percentage of uninteresting readings with Epsilon, on patients 221 and 230. The figure shows that as the value of $\epsilon$ increases (and correspondingly energy consumption gets reduced), NMSE also increases. Thus, the best value of $\epsilon$ is when NMSE is as low as possible with a large percentage of uninteresting readings. This happens at a value of Epsilon being roughly 0.2.

$$\text{MSE} = \frac{1}{n}\sum_{i=1}^{n}\left(Y_i - \hat{Y}_i\right)^2 \qquad (10)$$

Where, $n$ = number of data points, $Y_i$ = actual values, and $\hat{Y}_i$ = predicted values.

### 4.3 Tier-2 Assessment Results

The assessment at Tier-2 comprises the detection of anomalies in health parameter values streaming from various sensors to the LPU. The health parameter values streaming in at a particular time interval are aggregated to a single data vector and these are collected in a sliding window of size $\omega$ = 1024. The sliding window enables assessment of the streaming data as sequential blocks of size $\omega$ for identifying anomalies. Details on the approach to anomaly detection are included in Section 3.2. These anomalies represent health conditions that are not normal and are indications to raise an alarm. As elaborated in Section 3.2, *ntree* number of *iTrees* are created using the $\omega$ data points in the sliding window. We use a value of *ntree* = 100 in our experiments. Of these 100 *iTrees*, *ktree* = 20 number of *iTrees* are removed and replaced when a new set of $\omega$ data elements get collected in the sliding window. The values of *ntree* and *ktree* as 100 and 20 respectively are chosen at random. We experimented with other values of *ntree* and *ktree* as well and the results were similar.

*4.3.1 Anomaly Detection.* The proposed approach is assessed for its efficacy of anomaly detection using six health parameter (mentioned earlier) values of patients 221 and 230, namely: 1) Heart Rate; 2) Systolic Blood Pressure; 3) Diastolic Blood Pressure; 4) Pulse rate; 5) Respiratory rate; and 6) Oxygen saturation (SpO2). Figures 10 and 11 show through red bars the anomalies detected by the approach at the LPU. These bars correspond to when an alarm should be raised indicating that the patient is unwell and needs medical attention.

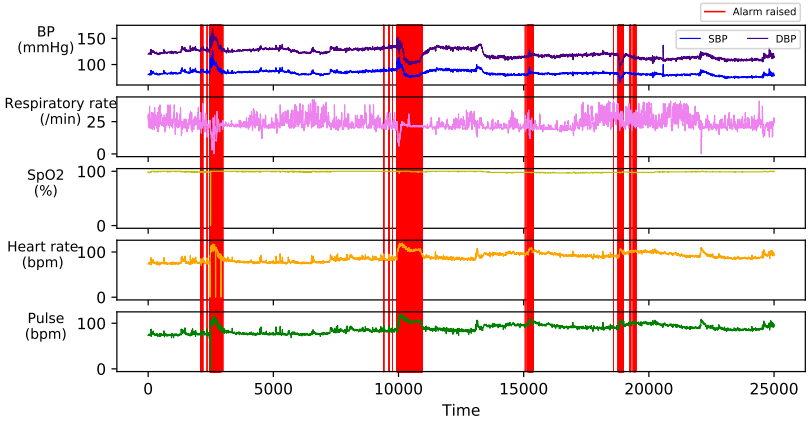

Fig. 10. Alarm Raised for patient 221

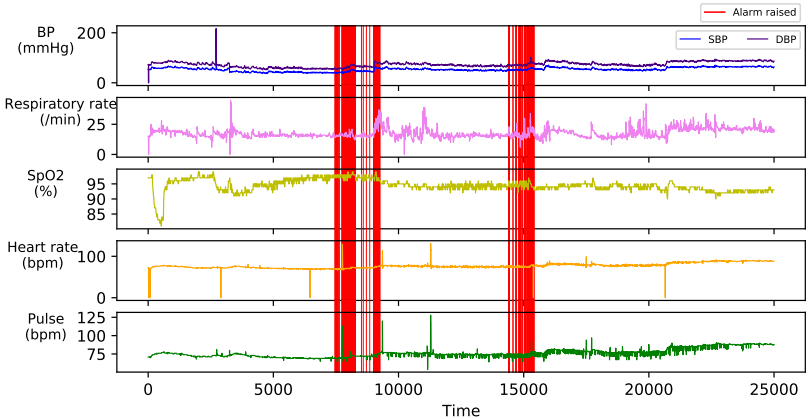

Fig. 11. Alarm Raised for patient 230

The effectiveness of the proposed approach in terms of precision, recall, and f1 score are shown in Table 3 for patients 221, 230, 237, 291, 401, and 442. The precision, recall, and f1 score are calculated as per the following equations

$$\text{Precision} = \frac{\text{TP}}{\text{TP+FP}} \quad (11)$$

Table 3. Precision-Recall Metrics

| Patient id | | precision | recall | f1-score | Patient id | | precision | recall | f1-score |
|---|---|---|---|---|---|---|---|---|---|
| **221** | 0 | 1.00 | 0.99 | 0.99 | **291** | 0 | 1.00 | 1.00 | 1.00 |
| | 1 | 0.72 | 0.99 | 0.84 | | 1 | 0.85 | 0.99 | 0.91 |
| | avg / total | 0.99 | 0.99 | 0.99 | | avg / total | 1.00 | 1.00 | 1.00 |
| **230** | 0 | 1.00 | 1.00 | 1.00 | **401** | 0 | 1.00 | 0.99 | 0.99 |
| | 1 | 0.92 | 0.99 | 0.95 | | 1 | 0.81 | 0.99 | 0.89 |
| | avg / total | 1.00 | 1.00 | 1.00 | | avg / total | 0.99 | 0.99 | 0.99 |
| **237** | 0 | 1.00 | 1.00 | 1.00 | **442** | 0 | 1.00 | 1.00 | 1.00 |
| | 1 | 0.87 | 0.99 | 0.92 | | 1 | 0.94 | 0.99 | 0.96 |
| | avg / total | 1.00 | 1.00 | 1.00 | | avg / total | 1.00 | 1.00 | 1.00 |

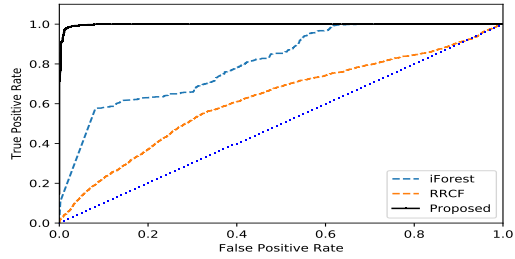

Fig. 12. Comparison with existing anomaly detection techniques

$$\text{Recall} = \frac{\text{TP}}{\text{TP+FN}} \quad (12)$$

$$\text{F1-score} = 2 \times \frac{(Precision * Recall)}{(Precision + Recall)} \quad (13)$$

Where, TP = True Positives, FP = False Positives, FN = False Negatives.

The detection accuracy for Patient 221 is 99.7% and 98.8% for patient 230 and is an indication of the potency of the approach.

### 4.3.2 Comparison with other Anomaly Detection Techniques.
We compare the proposed approach in terms of anomaly detection capabilities with two other standard anomaly detection techniques: *iForest* [36] and the Robust Random Cut Forest (RRCF) based anomaly detection technique [21]. We compare specifically with these two as they are both tree-based anomaly detection techniques and facilitate a fair comparison. We compare the three techniques in terms of Area under the ROC curve [22], a standard means for comparing such techniques. The ROC curve (short for Receiver Operator Characteristics curve) is a plot between the True Positive Rate (TPR) and False Positive Rate (FPR) for different threshold values for the assessed model. A large Area Under the ROC Curve (AUC) is a strong indicator of the efficacy of the model. The TPR and FPR are calculated as per the following equations

$$\text{TPR} = \frac{\text{TP}}{\text{TP+FN}} \quad (14)$$

$$\text{FPR} = \frac{\text{FP}}{\text{TN+FP}} \quad (15)$$

The ROC curves for the proposed approach, the *iForest*, and the RRCF are shown in Fig. 12. The AUC under the ROC characteristics for the proposed approach (attains 100% TPR very quickly, at an FPR of about 7.1%) clearly indicates a marked superiority in terms of anomaly detection efficacy when compared to the two standard and widely accepted techniques.

Moreover, the proposed approach is compared with other healthcare anomaly detection techniques in literature. We compare the approach with those in [23, 50] on the basis of False Positive rate (FPR) on patient 221 in the MIMIC Dataset. Our approach attains 100% True Positive rate (TPR) very quickly, at an FPR of about 7.1%. Whereas, Haque *et al.* [23] that use a Sequential Minimal Optimization (SMO) based anomaly detection reach 100% TPR at an FPR of about 24%. Smrithy *et al.* [50] who use a WMA based anomaly detection approach reach 100% TPR at an FPR of about 17%. This clearly demonstrates the superiority of the proposed approach in terms of anomaly detection efficacy.

We also compare the proposed approach with SVM [46] and J48 [12]. SVM achieves 100% TPR at an FPR of about 8%, whereas J48 achieves 100% TPR at an FPR of about 13%. It is important to note that the SVM and J48 approaches have other limitations: both approaches require a labeled training dataset which is frequently distorted and often unavailable; neither approach is compatible with real-time streaming and requires data to be stored in memory before processing. The proposed approach, therefore, clearly outperforms competing algorithms and enables efficient anomaly detection whilst consuming less energy and handling streaming data effectively.

### 4.4 Hardware Simulation of Tier-1 assessment

The validation of the algorithm at Tier-1 is incomplete without ensuring that it is indeed capable of working within the resource-constrained environments of a WBAN node. To demonstrate this, we simulate the hardware of a standard WBAN node and implement the proposed Tier-1 algorithm for detecting *faulty* and *uninteresting* health parameter values. A typical WBAN architecture comprising narrow-band wireless platforms such as Crossbow's Mica nodes and Texas Instruments' CC1010 and CC2400 [7, 30] based on ZigBee or Bluetooth wireless modules, is simulated. The simulation is done over MATLAB (Simulink) incorporating logic gates and MATLAB function blocks as shown in Figure 13. The working of the algorithm in the simulated environment is tested with data corresponding to the Heart Rate attribute of patient 221 from the MIMIC-1 dataset. Figure 14 shows the data-points of the Heart Rate attribute of patient 221 after assessment at Tier-1. Empty spaces in the graph indicate *faulty* or *uninteresting* readings that are eliminatated. Figure 15 shows the interpretation of the signal at the LPU. Empty spaces are replaced by values of the data immediately preceding the eliminated data points.

This is along expected lines and validates the practicability of the proposed approach in the resource-constrained environments of a WBAN node. The simulation ensures that the algorithm can easily be implemented on actual hardware.

### 4.5 Prototypical Implementation of the Proposed System

We assess the practicability of the proposed system in the real world through a prototypical implementation of a WBAN node. The node comprises an AD8232 Electrocardiogram (ECG) Monitor Sensor and an Arduino UNO microcontroller (a commercial implementation of a similar set-up would be significantly miniaturised and appropriately embedded in/on the body). The data on the heart's rate and rhythm continuously monitored by the sensor is assessed by the algorithm at the node. The data collected by the ECG sensor is in *ms*. Figure 16 shows a picture of the set-up on one of the authors to simulate the assessment at Tier-1.

The data monitored at the sensor is shown in Figure 17, and comprises 4500 data points. The Tier-1 algorithm implemented at the node eliminates 3794 data points that are deemed 'uninterested'. This works out to 84.31% of the data and would result in significant savings in term of energy expended on transmission. Figure 18 shows the graph with the reduced data points. Readings that are not far removed from immediately preceding data points are eliminated. It is clear from Figures 17 and 18 that the system and is able to effectively capture the nature of the health parameters in spite of the reduced data points. Subsequently, during Tier-2 assessment, no substantial anomalies were recorded, and an *AUC-ROC* score of 0.99 was recorded.

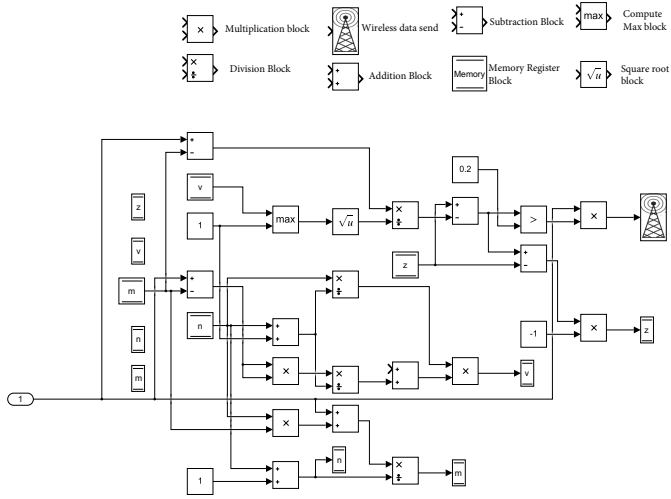

Fig. 13. Matlab block to simulate Tier-1 assessment

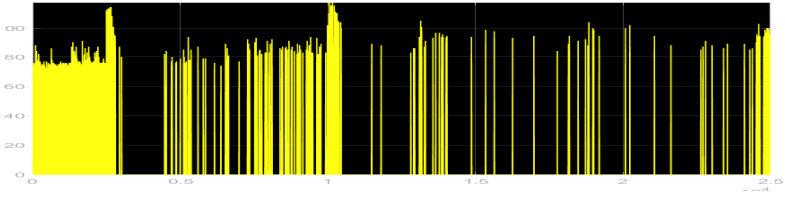

Fig. 14. Data points after Tier-1 Assessment

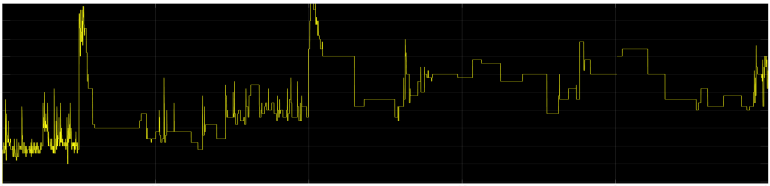

Fig. 15. Data interpretation at the LPU

A typical WBAN sensor, Mica2 [17], is used to calculate the power consumption in the proposed set-up. Polastre *et al.* [41] provide the total power consumption of a Mica2 sensor for various operations in Table 4. The total energy (in Joules) for each operation is calculated using Equation 16, assuming a constant voltage of 3 Volts [17].

$$Energy = Current(I) * Time(T) * Voltage(V) \qquad (16)$$

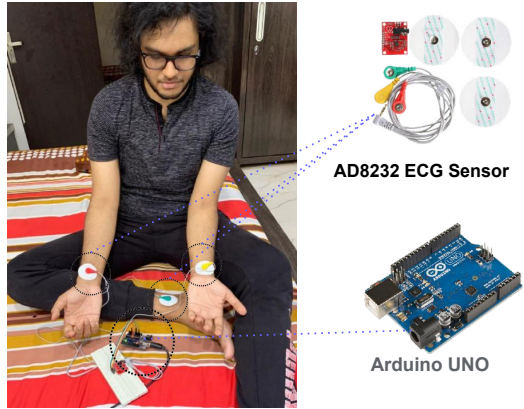

Fig. 16. Arduino Set-up and dataset generation on a set-up experimented on author

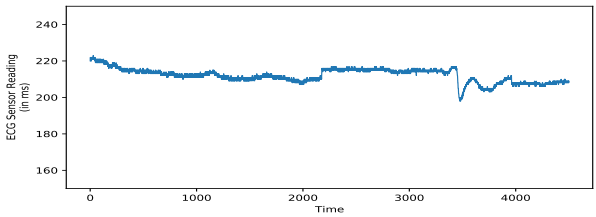

Fig. 17. Data on the heart's rhythm monitored by the ECG sensor

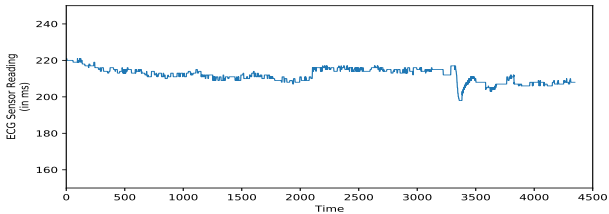

Fig. 18. Reduced data post Tier-1 assessment

From Table 4, we can compute that transmitting 1 byte of data requires: $20 \times 416 \times 10^{-6} \times 3 = 24.96$ $\mu$J of energy. Here, we demonstrate the energy efficiency of the proposed approach through the example of the respirator sensor (RESP) for patient 221 (Mimic dataset) in Table 2. The table shows that there are 25,000 data points generated, and assuming each data point to be of 4 bytes, Mica2 would require: $25000 \times 4 \times 24.96 \times 10^{-6} = 2.496$ J for transmission. This is the energy expended in the absence of Tier-1 assessment and all the sensor readings are transmitted to the LPU.

Our proposed approach does certain computations at the sensor and the energy expended for the computations also needs to be accounted for. From [11] we infer that computing each instruction requires 2.15 $n$J on a Mica2 sensor. The total energy spent on computations is calculated using

Table 4. Mica2 Sensor power consumption on each operation

| Operation | Time (s) | I (mA) |
|---|---|---|
| Initialize radio | $350E-6$ | 6 |
| Turn on radio | $1.5E-3$ | 1 |
| Switch to RX/TX | $250E-6$ | 15 |
| Time to sample radio | $350E-6$ | 15 |
| Evaluate radio sample | $100E-6$ | 6 |
| Receive 1 byte | $416E-6$ | 15 |
| Transmit 1 byte | $416E-6$ | 20 |
| Sample sensors | 1.1 | 20 |

Equation 17.

$$Computation\ energy = Number\ of\ Instructions \times 2.15 \times 10^{-9} J \quad (17)$$

The total number of instructions computed at Tier 1 for the respiratory sensor for patient 221 is obtained by converting the assessment step into assembly code (symbolic machine code), and works out to 1,850,081 instructions. The total energy expended for computations, therefore is: $1850081 \times 2.15 \times 10^{-9}$ = 0.004 J. The energy expended in transmission of data to the LPU works out to 32.1% of that expended to transmit all the datapoints which is 0.801 J (32.1% of 2.496 J). This is because Table 2 shows that the proposed method discards 67.9% of the respiratory sensor data generated for patient 221. The total energy expended is therefore: 0.004 J (computation energy) + 0.801 J (transmission energy) = 0.805 J. This works out to a saving of 67% (0.805 J as compared to 2.496 J) of energy owing to the use of the proposed 2-Tier assessment approach and is quite significant.

## 5 CONCLUSION

An approach to eliminate 'uninteresting' health parameter measurements by sensors in Wireless Body Area Networks (WBAN) was proposed in this paper. The idea involves: 1) eliminating measurements that have not deviated much from their immediately preceding values and hence provide no new information; and 2) elimination of measurements that are significantly deviated from the normal and are most likely faults. Removal of such uninteresting measurements markedly reduces the energy required to transmit measured health parameter values from the sensors to the Local Processing Unit (LPU). In our experiments, the elimination of data at the site of the sensors was around 90% and indicated a proportional degree of energy savings. Such computations at the resource-constrained senor nodes are usually viewed suspiciously as being impracticable in real environments. To demonstrate the feasibility of the proposed approach, we conducted a hardware simulation of a typical sensor node in a WBAN and implemented and executed the algorithm over it. Results of the simulation were in conformance with our expectations demonstrating the practical viability of the approach.

The second tier of assessment involved harnessing an anomaly detection model that is capable of handling streaming data and working within the confines of the relatively resource-constrained environment of an LPU, usually a mobile device, as compared to the infinitely capable cloud. Detecting anomalies at the LPU instead of the cloud leads to significant energy savings otherwise expended on data transmission to the cloud and also minimises latency. The anomaly detection enables identifying anomalous readings that indicate a possible medical condition and sometimes even an emergency. The efficacy of the anomaly detection model was compared with standard models and shown to be efficacious.

The results are promising and we aim to combine the proposed approach with a parallel endeavour of ours wherein the various sensors in a WBAN more effectively co-exist and communicate with the LPU through superior spectrum utilisation.